\documentclass[journal=jacsat,manuscript=article]{achemso}

\usepackage[version=3]{mhchem} 

\usepackage{soul}
\usepackage{amsmath,bm}
\usepackage{siunitx}
\usepackage{braket}

\author{Armando Pezo}
\affiliation{Laboratoire Albert Fert, CNRS, Thales, Université Paris-Saclay, 91767, Palaiseau, France}
\affiliation{Aix-Marseille Université , CNRS, CINaM, Marseille, France}
\email{armando-arquimedes.pezo-lopez@cnrs-thales.fr}

\author{Andrés Saul}
\affiliation{Aix-Marseille Université , CNRS, CINaM, Marseille, France}

\author{Aurélien Manchon}
\affiliation{Aix-Marseille Université , CNRS, CINaM, Marseille, France}

\author{Rémi Arras} 
\affiliation{CEMES, Université de Toulouse, CNRS, 29 rue Jeanne Marvig, F-31055, Toulouse, France}
\email{remi.arras@cemes.fr}

\title[An \textsf{achemso} demo]
  {Ferroelectric control of spin and orbital Rashba effects at the Ni/HfO$_2$ interface}

\abbreviations{IR,NMR,UV}

\begin{document}

\begin{abstract}
We predict the giant ferroelectric control of interfacial properties of Ni/HfO$_2$, namely, (i) the magnetocrystalline anisotropy and (ii) the inverse spin and orbital Rashba effects. The reversible control of magnetic properties using electric gating is a promising route to low-energy consumption magnetic devices, including memories and logic gates. Synthetic multiferroics, composed of a ferroelectric in proximity to a magnet, stand out as a promising platform for such devices. Using a combination of $ab$ $initio$ simulations and transport calculations, we demonstrate that reversing the electric polarization modulates the interface magnetocrystalline anisotropy from in-plane to out-of-plane. This modulation compares favorably with recent reports obtained upon electromigration induced by ionic gating. In addition, we find that the current-driven spin and orbital densities at the interface can be modulated by about 50\% and 30\%, respectively. This giant modulation of the spin-charge and orbit-charge conversion efficiencies opens appealing avenues for voltage-controlled spin- and orbitronics devices. 
\end{abstract}


The design of extrinsic multiferroics, formed by the association of a ferroelectric material with a ferromagnet in heterostructures, has been a subject of intense research aiming to obtain large magnetoelectric couplings~\cite{Garcia:2015}. Indeed, switching the magnetization by applying an electric field represents an extraordinary opportunity to reduce the power consumption of magnetic memory devices compared to using large magnetic fields or spin-transfer torques (STT). For a few years now, much effort has been devoted to manipulating the magnetization through spin-orbit torques (SOT), which takes advantage of spin currents or densities generated electrically through spin-charge interconversion, i.e., the spin Hall or Rashba-Edelstein effects. These new approaches have led to the emergence of novel logic and memory devices such as the magneto-electric spin-orbit (MESO)~\cite{Manipatruni:2019}, the ferroelectric spin-orbit (FESO)~\cite{Noel:2020} or the SOT-magnetic random access memory (SOT-MRAM)~\cite{Manchon:2019} devices.  

In these devices, spin-orbit coupling plays a central role as it governs the spin-charge interconversion phenomena. Ferroelectric compounds can substantially boost these mechanisms by altering the crystal structure in a non-volatile manner. Consequently, non-volatile switching of the spin texture and spin-charge interconversion can be achieved by reversing the ferroelectric polarization using electrical gates~\cite{Mirhosseini:2010,Noel:2020,Fang:2020,Varotto:2021}. The structural change accompanying ferroelectric reversal is also expected to alter the orbital transport that has recently dragged attention as a non-relativistic (spin-orbit coupling-free) alternative to spin transport~\cite{Dongwook:2021, Krishnia:2024, Pezo:2022, Pezo:2023, Johansson:2021}. Therefore, extrinsic multiferroics demonstrating efficient control of the spin and orbital properties are highly desired for low-power consumption devices.

However, the use of ferroelectric compounds in miniaturized devices faces several challenges~\cite{Ma:2002}. The electric polarization is indeed known to vanish below a critical thickness when the ferroelectric material is grown in thin films because of the increase of the depolarizing field. A second important problem comes from the fact that the most widely used ferroelectric compounds so far are oxides with a perovskite structure that is difficult to grow on silicon substrates, making them incompatible with CMOS technologies. Of these observations, the recent discovery of ferroelectricity in doped hafnia (HfO$_2$) films~\cite{Boscke:2011, Park:2015, Park:2017, Yun:2022}, which do not suffer from the aforementioned issues, has been welcomed with enthusiasm for the design of ferroelectric-based applications or memories involving resistive-switching processes~\cite{Mikolajick:2018, Banerjee:2022}. 

In spintronic devices, hafnia needs to be incorporated in thin-film heterostructures in which it is interfaced with ferromagnetic metallic electrodes~\cite{Banerjee:2022}. Several studies have already demonstrated the possibility of controlling the magnetic properties (spin polarization~\cite{Wei:2019}, magnetic anisotropy~\cite{Vermeulen:2019}, magnetization~\cite{Yang:2019, Mikheev:2022, He:2023}) by reversing the electric polarization in hafnia thin films. In this Letter, using first-principles calculations, we assess the ferroelectric control of spin-orbit properties of Ni/HfO$_2$ bilayer. We demonstrate that switching the electric polarization not only tunes interfacial magnetism but also modifies the interfacial magnetocrystalline anisotropy from in-plane to out-of-plane. In addition, we show that current-driven spin and orbital densities can be controlled in a non-volatile manner by ferroelectricity, opening avenues for the gate-controlled spin-charge and orbital-charge interconversion.

\begin{figure}[!htb]
    \centering
    \includegraphics[width=.85\linewidth]{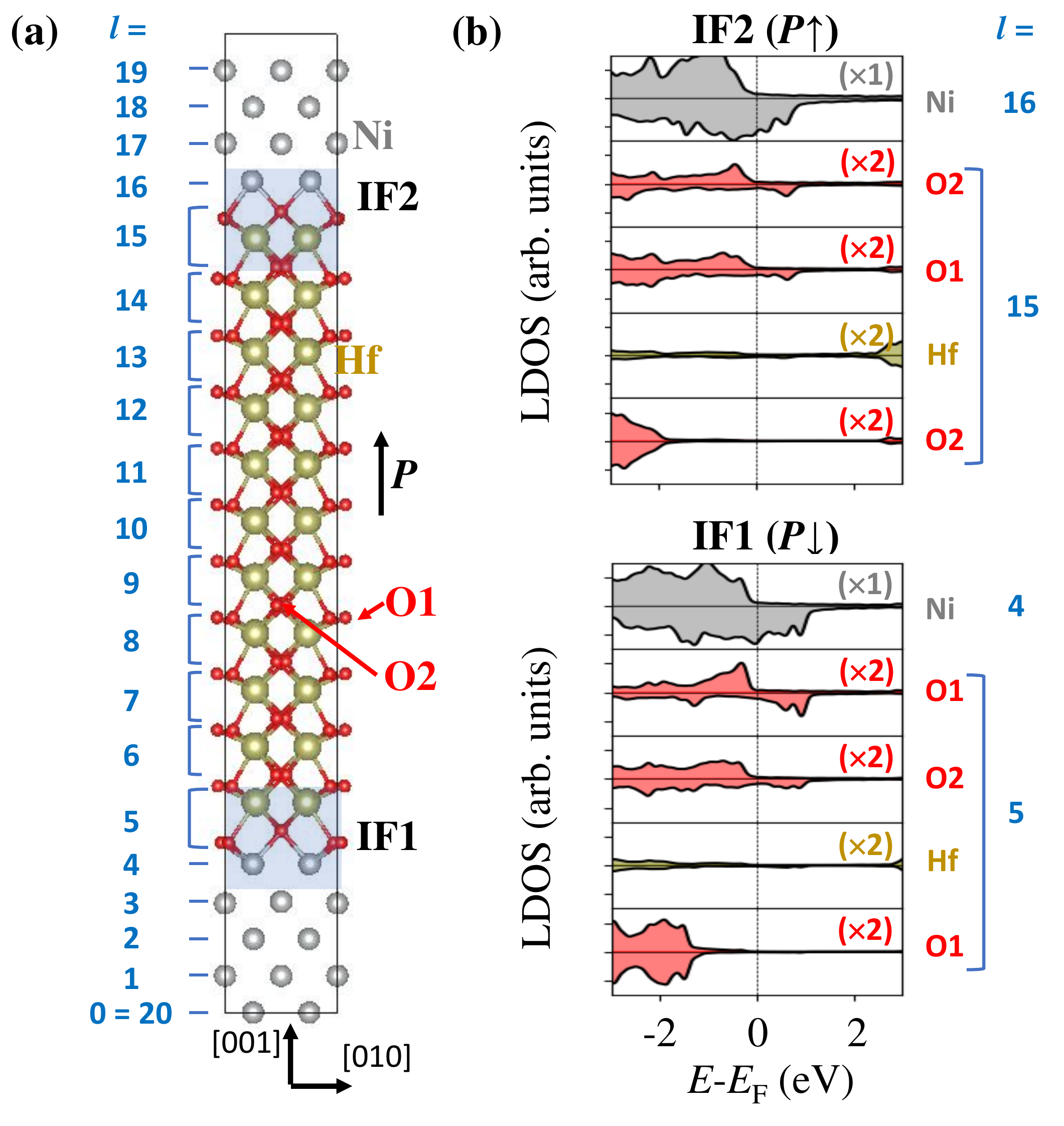}
    \caption{(a) Atomic structure of the Ni/HfO$_2$ superlattice used for the calculations, (b) Layer-resolved density of states (LDOS) calculated for the two interfaces IF1 and IF2. Positive and negative DOS correspond to the majority/minority-spin channels, respectively. The LDOSs of HfO$_2$   atomic layers have been multiplied by 2 for better clarity. \label{fig1}}
\end{figure}

To model the effect of the electric polarization on the Ni electrode, we considered a periodic Ni/HfO$_2$ superlattice, shown in Fig.~\ref{fig1}(a) and formed by a total of 20 atomic layers and two interfaces at which the electric polarization $P$ is either pointing inward (P$\downarrow$, IF1) or outward (P$\uparrow$, IF2). Our numerical investigation is based on the density functional theory (DFT) and nonequilibrium transport calculations. More details about the numerical methodology and the optimization of the structure are provided in the section S.I. of the supplementary materials.

As can be seen in Fig.~\ref{fig1}(a), HfO$_2$ possesses two inequivalent oxygen atoms, labeled O1 and O2. At the center of the film, the positions of the O1 atoms are off-centered, with a difference of out-of-plane $z$ coordinates between the Hf and O atoms of approximately 27.4~pm. This observation confirms that the electric polarization is stable in the HfO$_2$ film, although slightly reduced with respect to the bulk value ($\simeq 28.5$~pm) due to the presence of a depolarizing field. At the two interfaces, we observe a variation of this difference of $z$ coordinates, which increases up to 27.8~pm at IF1, while it tends to reduce to 25.1~pm at IF2 [see Fig.~S1(c)]. As shown in Fig.~S1(d), the electric polarization is accompanied by an internal electric field of approximately $1.73 \times 10^8$~V~m$^{-1}$ that shifts the layer-resolved densities of states (LDOS) higher in energy when moving from layer $l = 6$ to $l = 14$. In Fig.~\ref{fig1}(b), we detail the contributions to the LDOS of the different atoms forming the two interface layers. The electric polarization and the different relaxation of the O atoms at the interfaces IF1 and IF2 result in a difference of charge transfer and hybridization between the O-$p$ and Ni-$d$ orbitals, as confirmed by the variation of the atomic spin moments and Mulliken charges reported in Table.~S1. Because we choose to model both interfaces between the Ni electrode and with a full oxygen layer composed of two O1 and two O2 atoms, the whole HfO$_2$ film is not stoichiometric by construction, which explains the shift of the Fermi level toward lower energies. The LDOSs show that the Ni bands at IF1 are more depopulated than at IF2, and that their hybridization with the orbitals of O1 is stronger than with O2 at this interface, while it is almost equivalent for the two types of oxygen atoms at IF2. This difference of charge transfer between the two interfaces is responsible for a variation of the local magnetic moment of 1.44~$\mu_\textrm{B}$ per interface area (see Table~S1), characteristic of the emergence of a magnetoelectric coupling, as discussed in Sec. S.II.B. The data displayed in Fig.~S2 shows that the variation of charge transfer and local spin magnetic moment is not linear as a function of $P$; as explained below, these properties mostly depend on the position of the O1 atoms located at the interface.

\begin{figure}[!htb]
    \centering
    \includegraphics[width=.85\linewidth]{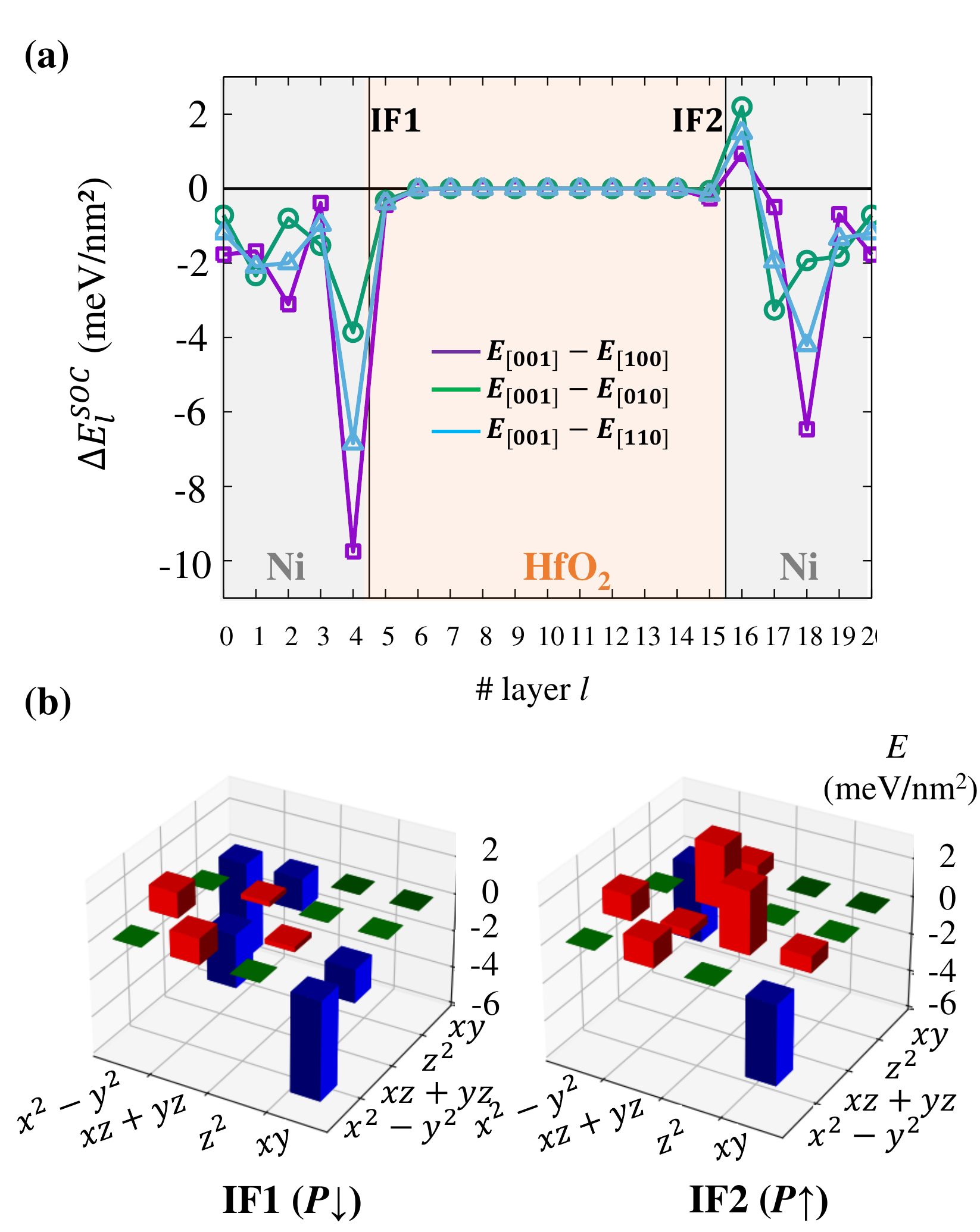}
    \caption{(a) Layer-resolved MCAE $\Delta E^\textrm{SOC}_l$ calculated for different crystallographic directions, (b) $d$-orbital contributions to the MCAE of the Ni atoms in the interface layer IF1 ($l = 4$ - P$\downarrow$) and IF2 ($l = 16$ - P$\uparrow$). The red and blue bars correspond to positive and negative energies. \label{fig2}}
\end{figure}

Because of the difference of orbital overlaps between the two interfaces, i.e., when $P$ is reversed, it is also expected that the magnetocrystalline anisotropy energy (MCAE) varies upon switching the electric polarization. We find that, for the whole supercell, the easy axis is oriented along the [001] direction, with a total MCAE of $-13.6$, $-23.7$ and $-18.9$~meV~nm$^{-2}$ with respect to the [010], [100] and [110] axes, respectively. In Fig.~\ref{fig2}(a), we report the layer projection of the MCAE, $\Delta E^\textrm{SOC}_l$, in order to distinguish the individual contribution of each Ni/HfO$_2$ interface. Whereas, in the center of the Ni layer, $\Delta E^\textrm{SOC}_l$ remains slightly negative ($\simeq -1.3$~meV~nm$^{-2}$ with the [110] direction, close from the value of $-1.5$~meV~nm$^{-2}$ calculated for strained bulk Ni), it displays strong variations in the two Ni atomic layers at the interface, with a decrease down to $\Delta E^\textrm{SOC}_4 = -6.8$~meV~nm$^{-2}$ at IF1 ($l =4$), and an increase up to the positive value of $\Delta E^\textrm{SOC}_{16}=+1.4$~meV~nm$^{-2}$ at IF2 ($l =16$). At IF1, we thus predict that the perpendicular magnetic anisotropy (PMA) is reinforced, whereas it is weakened at IF2, enabling a net in-plane magnetic anisotropy. To assess the overall impact of ferroelectricity on the MCAE, we define the anisotropy modulation as $\Delta E^\textrm{SOC}_P=\sum_{i=1}^4\Delta E^\textrm{SOC}_i-\sum_{i=16}^{20}\Delta E^\textrm{SOC}_i$. We find that reversing the ferroelectric polarization results in a change of MCAE of about $\Delta E^\textrm{SOC}_P=5.6$~meV~nm$^{-2}$ (8~meV~nm$^{-2}$) for the [110] ([100]) direction. This modification is remarkably large as it is comparable to the PMA of Fe/MgO interfaces ($\approx 8.5$~meV~nm$^{-2}$)~\cite{Yang:2011} and is of the same order as the voltage-controlled MCAE change reported by Bauer et al.~\cite{Bauer:2015}. This efficient ferroelectric tuning of the magnetic anisotropy opens inspiring perspectives for low-energy control of magnetism. 

This strong variation of the MCAE stems from the dramatic impact of the ferroelectricity on the hybridization between Ni-$3d$ and O-$2p$ orbitals (see Fig.~S2 and~S3~\cite{SI}) but also on the $d$-$d$ interactions through lattice distortions and charge transfer. From Fig.~\ref{fig2}(b), we can indeed see that the strong contribution toward the PMA ($\Delta E^\textrm{SOC}_l < 0$) in the IF1($l=4$) layer is mostly due to interactions between $d_{xy}$ and $d_{x^2-y^2}$ orbitals. At IF2($l=16$), the magnitude of the interaction between the $d_{xy}$ and $d_{x^2-y^2}$ orbitals is decreased, when compared with those calculated at IF1, and we observe at the same time an increase of the positive interactions between $d_{z^2}$ and $d_{xz},d_{yz}$ orbitals, which become dominant. According to the projected DOS displayed in Fig.~S3(b), the states for which these orbital contributions are mainly formed strongly hybridize with O-$(p_x+p_y)$ orbitals.

\begin{figure}[!ht]
    \centering
    \includegraphics[width=\linewidth]{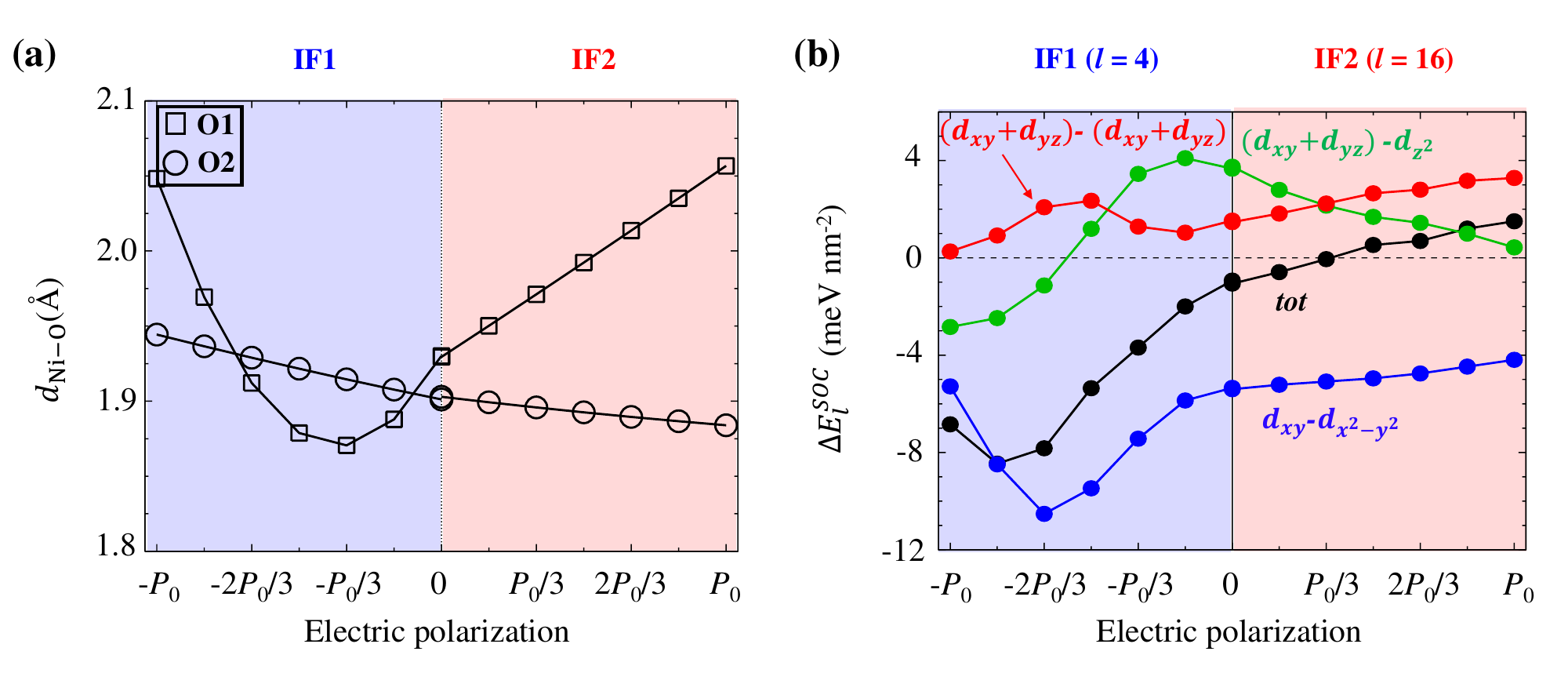}
    \caption{Variation of the (a) chemical bond lengths $d$ between the nearest Ni and O atoms and (b) MCAE projected on the interface Ni layers, with the corresponding main $d$-$d$ contributions. \label{fig3}}
\end{figure}

To describe how the MCAE varies with the Ni-O interatomic distances at the interface, we compute these interface properties as a function of the electric polarization. To do so, as explained in details in Sec.~S.I.B~\cite{SI}, we compute the MCAE for different structures generated by linearly interpolating the atomic positions between a supercell with a centrosymmetric HfO$_2$ film, in which the atomic positions have been kept fixed inside the hafnia film and relaxed at the interface and in the Ni layer, and the fully-optimized polar structure. From the structural optimization, we found that the linear variation of $P$ from $-P_0$ to $+P_0$ modeled with this method also induces a linear variation of the averaged out-of-plane Ni-O1 and Ni-O2 distances at the Ni/HfO$_2$ interface (not shown). As illustrated in Fig.~\ref{fig3}(a), if the variation of the averaged Ni-O chemical bond lengths remains linear with O2 atoms in the whole range of polarization, we observe a non-linear variation with the polar O1 atoms when the polarization is negative, which is a consequence of an additional relative displacement of the Ni/O atoms along the $[100]$ axis. With this in-plane displacement, the interface Ni atoms that are initially located in a bridge position, i.e., with two oxygen atoms as first neighbors, move in a hollow position and get closer to a third oxygen atom at the interface IF1. The averaged Ni-O bond length then passes by a minimum of $d_\textrm{Ni-O1} = 1.87$~\AA{} for $P = -P_0/3$ and reaches two maxima of $d_\textrm{Ni-O1} = 2.05$~\AA{} at $P = -P_0$ and  $d_\textrm{Ni-O1} = 2.06$~\AA{} at $P = +P_0$. The Ni-O distances at the interfaces correlates with the variations of the local MCAE, as shown in  Fig.~\ref{fig3}(b). Similarly, the MCAE increases linearly between $0$ and $+P_0$ and is non-linear for negative electric polarization at IF1. The change of $\Delta E^\textrm{SOC}_l$ is mostly dominated by the change of the negative interaction between $d_{xy}$ and $d_{x^2-y^2}$ orbitals and positive interaction between $d_{xz}$ and $d_{yz}$ orbitals. The decrease of these interactions when $P$ decreases favors the stabilization of the PMA, but is accompanied by an increase of the $(d_{xz},d_{yz})$-$d_{z^2}$ interaction, which has the opposite effect. 

Besides the ferroelectric control of the MCAE, reversing the ferroelectric polarization is also expected to massively impact the spin and orbital textures at the interface, which is of direct interest to spin-charge and orbital-charge interconversion \cite{Noel:2020,Hamdi:2023}. As a matter of fact, the inversion symmetry breaking at the interface promotes an orbital texture that is odd in momentum (see Fig.~S5~\cite{SI}). In the presence of spin-orbit coupling, this orbital texture is accompanied by a spin texture~\cite{Bihlmayer:2022} that unlock the so-called inverse spin~\cite{Edelstein:1990} and orbital~\cite{Go:2017} Edelstein effects, i.e., the generation of a spin and orbital density induced by a charge current. Switching the ferroelectric polarization is expected to modify the orbital hybridization between O1 and Ni at the interface and thereby tune both inverse spin and orbital Edelstein effects. We compute the nonequilibrium spin and orbital densities, $\delta\langle S_y\rangle$ and $\delta\langle L_y\rangle$, using the Kubo formula within the linear-response theory~\cite{bonbien_kubo} (more details are given in the supplementary information~\cite{SI}). Since the noncentrosymmetric spin and orbital textures are well localized at the interface, as confirmed by Fig.~S4~\cite{SI}, we focus our attention on the contributions coming from Ni and O1 atoms at IF1 and IF2. Because the calculated densities depend on the energy of the Fermi level, we also restrict our study to an energy window delimited by the band gap of HfO$_2$ in the vicinity of the two interfaces, i.e., between $-1.8$($-1.5$)~eV and $+2.5$($+2.9$)~eV at IF1(IF2), where only the bands of Ni and O participate to the transport properties (See Fig.~\ref{fig1}(b) and Fig.~S1~\cite{SI}).

\begin{figure}
\centering
\includegraphics[width=1\linewidth]{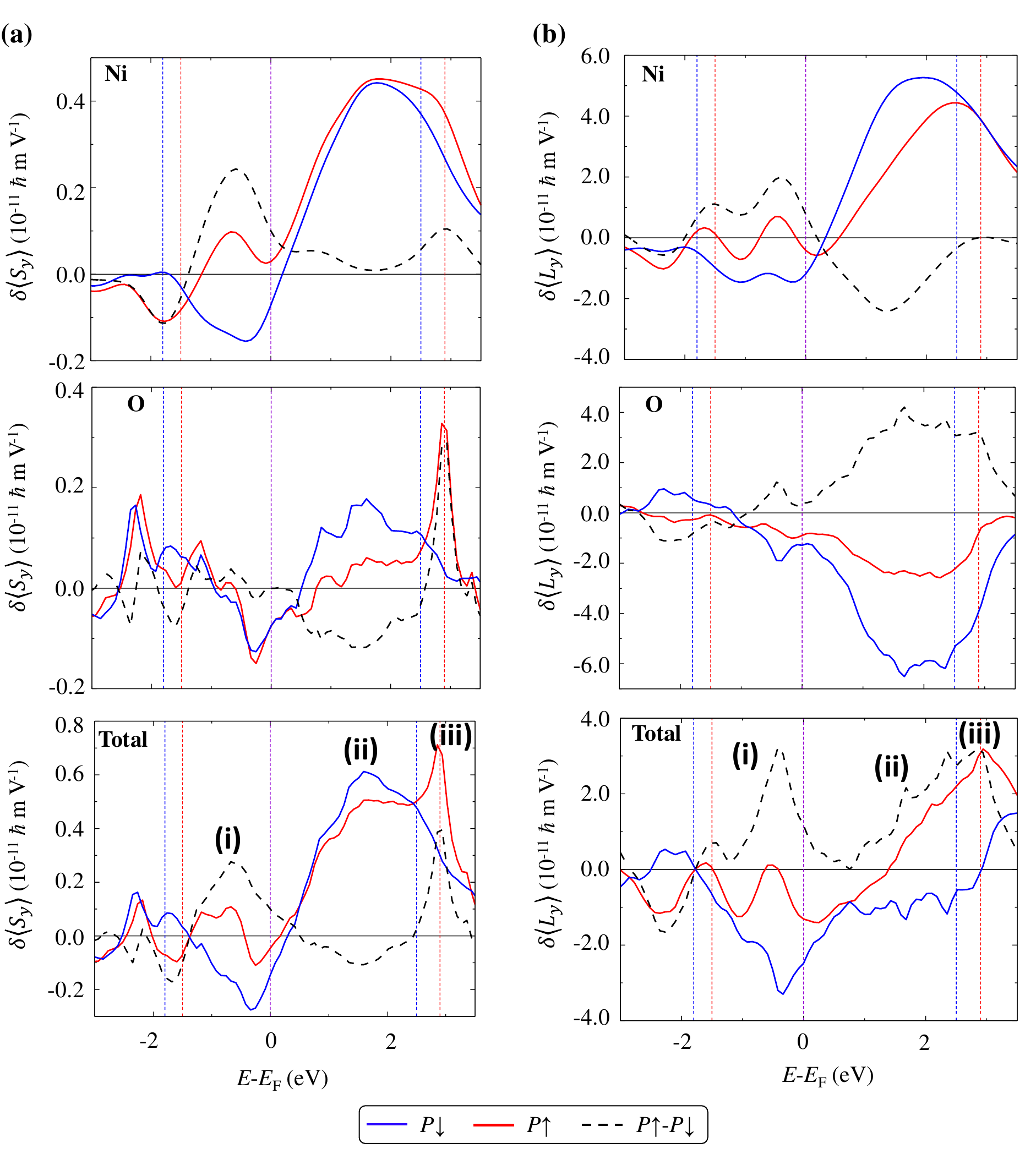}
\caption{Spin (a) and Orbital (b) responses to an in-plane electric field calculated for a top Ni electrode. The response for an outward(inward) electric polarization $P\uparrow$($P\downarrow$) is calculated considering the contribution of the interface IF2 (IF1, multiplied by $-1$). More details are given in Section S.I.D and S.III from the supplementary materials~\cite{SI}). The dashed black lines correspond to the difference of both responses, i.e., 2$\delta\langle S_y\rangle_P$ (a) and 2$\delta\langle L_y\rangle_P$. The red and blue dotted lines show the limits of the local HfO$_2$ band gap in the vicinity of the two interfaces. \label{fig4}}
\end{figure}

Figure~\ref{fig4} displays both spin and orbital densities, $\delta\langle S_y\rangle$ and $\delta\langle L_y\rangle$, for the electric polarization pointing outward (P$\uparrow$, red lines) and inward (P$\downarrow$, blue lines).
If the slab were perfectly symmetric, and if the electric polarization did not have any influence, $\delta\langle S_y\rangle$ (and $\delta\langle L_y\rangle$) would be equal for P$\uparrow$ and P$\downarrow$. The influence of the electric polarization can be analyzed by parsing the density into a symmetric, $\delta\langle S_y\rangle_0=(\delta\langle S_y\rangle_{P\uparrow}+\delta\langle S_y\rangle_{P\downarrow})/2$, and an antisymmetric part, $\delta\langle S_y\rangle_P=(\delta\langle S_y\rangle_{P\downarrow}-\delta\langle S_y\rangle_{P\uparrow})/2$, the latter being directly associated with the electric polarization. A similar definition applies to the orbital density $\delta\langle L_y\rangle$. In Table \ref{tab:IFacc}, we report the values of the symmetric and antisymmetric parts of the spin and orbital densities at Fermi level. We find that the average spin density located on the interfacial Ni is $\delta\langle S_y\rangle_0=-2\times10^{-13}\hbar$~m~V$^{-1}$, which, using a free electron model~\cite{Edelstein:1990} with a lattice constant of 2.5~\AA, corresponds to an effective Rashba strength of $\sim10^{-13}$~eV~m. Upon reversing the ferroelectric polarization, the modulation of the spin density amounts to $\delta\langle S_y\rangle_P=5\times10^{-13}\hbar$~m~V$^{-1}$, which corresponds to a giant modulation $\eta_S=\delta\langle S_y\rangle_0/\delta\langle S_y\rangle_P=-250\%$. Interestingly, the spin density at the Fermi level on the O1 atoms near the interface is unaffected by the polarization. Summing both contributions, the total spin density at the interface can be modulated by about $\eta_S\approx 53\%$. As a comparison, Mishra et al.~\cite{Mishra:2019} recently demonstrated that in Pt/Co($d$)/GdOx thin films, the Rashba field $H_R/j_e$ could be modulated from $-500$ to $+250$~Oe~A$^{-1}$~m$^2$ by voltage-driven oxygen migration. The corresponding modulation of the spin density can be obtained using $S=(dM_s/\Delta)H_R/j_e$~\cite{Manchon:2008}, where $d=0.8$ nm and $M_s=1400$~kA m$^{-1}$ are the thickness and saturation magnetization of the Co layer, and $\Delta$ is the exchange energy ($\approx2$ eV). Using these parameters, we find that the spin density is tuned between -10 and +5$\times10^{-13}\hbar$~m~V$^{-1}$, which compares favorably with our prediction.

\begin{table}[!ht]
\centering 
\begin{tabular}{ccccccc}               
\hline
\hline
 & $\delta\langle S_y\rangle_0$ & $\delta\langle S_y\rangle_P$ & $\eta_S$ (\%) & $\delta\langle L_y\rangle_0$ & $\delta\langle L_y\rangle_P$ &$\eta_L$ (\%)\\ 
Ni & $-0.02$ & 0.05 & $-250$  & $-0.96$ & 0.21 & $-22$ \\
O & $-0.074$ & 0 & 0 & $-1.09$ & 0.19 & $-17$ \\
Total & $-0.092$ & 0.05 & $-53$ & $-1.9$ & 0.6 & $-32$\\
\hline
\hline
\caption{Antisymmetric and symmetric parts of the spin and orbital densities at Fermi level. The unit is $10^{-11}\hbar$~m~V$^{-1}$.\label{tab:IFacc}} 
\end{tabular}
\end{table}

The orbital densities displayed in Fig.~\ref{fig4}(b) are about one order of magnitude larger than the spin density, in agreement with the literature~\cite{Dongwook:2021, Krishnia:2024, Santos:2023}. At Fermi level, we obtain an average interfacial orbital density of about $-1.9\times10^{-11}\hbar$~m~V$^{-1}$, which is comparable to the orbital densities obtained at transition metal interfaces~\cite{Salemi:2021} and one order of magnitude smaller than the giant orbital density found at Al/Co interfaces~\cite{Nikolaev2024}. The ferroelectric modulation of the orbital densities is about 32\%, which is remarkable.

Away from Fermi level, the spin and orbital densities display two extrema, labeled (i) and (ii) around energies of approximately $-0.6$~eV and $+1.7$~eV, which also correspond to a maximal modulation upon polarization reversal. For energies below $E_\textrm{F}$, the bands have a predominant Ni-$d$ character and the calculated spin and orbital responses are moderate. This means that although Ni-$d$ states provide reasonably large spin and orbital projections around the Fermi level, their higher localized nature limits their overall magnitude. This is in contrast with the densities observed around (ii), where $sp$ bands predominate, leading to enhanced responses, in agreement with the large value of the longitudinal conductivity $\sigma_{xx}$, as shown in Fig.~S6~\cite{SI}. We notice a narrowing of the spin density peak (ii), when switching from P$\downarrow$ to P$\uparrow$, and a shift in energy of the maximum of the orbital response, from $E_\textrm{F}+1.93$~eV to $E_\textrm{F}+2.47$~eV, which follows the shift of the electric potential and the response contribution of Ni atoms. A third maximum (iii) in the spin and orbital responses can also be noticed at $E \sim E_\textrm{F}+2.9$~eV; such an energy corresponds to the local conduction band minimum of HfO$_2$ for the P$\uparrow$ configuration (IF2 interface), where the oxygen atoms interact more strongly with Hf atoms due to their larger proximity in energy, enhancing the spin-orbit interaction and, consequently, the spin density. This peak is shifted toward higher energies and decreases in magnitude upon polarization switching. At $P=-P_0$, the peak completely disappears because of the too large Hf-O1 distance for the P$\downarrow$ configuration (IF1 interface - see Fig.~S7~\cite{SI}).

In summary, we have found that the spin and orbital properties at Ni/HfO$_2$ interface can be controlled through the ferroelectric polarization via the modulation of the Ni-O bond length. Remarkably, our calculations suggest that the MCAE can be switched from in-plane to out-of-plane upon reversing the electric polarization. At the same time, the spin and orbital textures at Fermi level are also substantially tuned, resulting in modulations of the spin-charge and orbital-charge interconversion efficiency of several tens of percent. This finding makes Ni/HfO$_2$ particularly appealing for voltage-controlled memory and logic devices.

\begin{acknowledgement}
This work was supported by the ANR ORION project, grant ANR-20-CE30-0022-01 of the French Agence Nationale de la Recherche, by the Excellence Initiative of Aix-Marseille Université-A*Midex, a French ”Investissements d’Avenir” program, by the France 2030 government grants managed by the French National Research Agency PEPR SPIN [SPINMAT] ANR-22-EXSP0007 and [SPINTHEORY] ANR-22-EXSP-0009 and by the EIC Pathfinder OPEN grant 101129641 “OBELIX”. This work was granted access to the HPC resources of CALMIP (Allocations No. 2022-2024/P19004 and P1229) and CINES (Allocation AD010915807R1).

\end{acknowledgement}


\providecommand{\latin}[1]{#1}
\makeatletter
\providecommand{\doi}
  {\begingroup\let\do\@makeother\dospecials
  \catcode`\{=1 \catcode`\}=2 \doi@aux}
\providecommand{\doi@aux}[1]{\endgroup\texttt{#1}}
\makeatother
\providecommand*\mcitethebibliography{\thebibliography}
\csname @ifundefined\endcsname{endmcitethebibliography}
  {\let\endmcitethebibliography\endthebibliography}{}

\setcounter{figure}{0}
\renewcommand{\figurename}{Fig.}
\renewcommand{\thefigure}{S\arabic{figure}}

\setcounter{figure}{0}
\renewcommand{\tablename}{Table}
\renewcommand{\thetable}{S\arabic{table}}

\renewcommand{\thesection}{S.\Roman{section}}

\section{Supplementary materials: Spin and Orbital Rashba effects at the Ni/HfO$_2$ interface}

\section{Details of the calculations}
Our numerical investigation is based on the density functional theory (DFT) and relies on two-step calculations consisting in (i) a structural relaxation of the supercells using the projector augmented-wave (PAW)~\cite{Blochl:1994,Kresse:1999} method implemented in the \textsc{VASP} code~\cite{Kresse:1996a,Kresse:1996b} and (ii) the subsequent evaluation of transport properties based on a localized basis set with the SIESTA code~\cite{siesta_method}. In both cases, we used the Perdew-Burke-Ernzerhof revised for solids (PBESol)~\cite{Perdew:2008,*Perdew:2009} exchange-correlation functional.

\subsection{Atomic structures and general parameters}

The Ni/HfO$_2$(001)-interface properties were modeled using supercells formed by the association of 11 atomic layers of HfO$_2$ and 9 atomic layers of Ni as shown in Fig.~1(a). With such structures which contain no Ni surfaces, but two different different Ni/HfO$_2$ interfaces, it is possible to study the local effect of the two possible directions of the electric polarization at the same time, the electric polarization being always parallel to the out-of-plane axis: Hence we label the first interface IF1, at which the electric polarization $\bm{P}$ is pointing inward (Ni $\rightarrow$ HfO$_2$) and the second interface IF2, where $\bm{P}$ is outward (HfO$_2 \rightarrow$ Ni). The in-plane lattice parameters have been fixed to the equilibrium values calculated for the bulk orthorhombic and ferroelectric phase of HfO$_2$ (space group 29-$Pca2_1$). The Ni face-centered cubic lattice is accommodated on HfO$_2$ according to the (001)[110] epitaxy relation; we calculated that HfO$_2$ induces an in-plane tensile strain on Ni of $-4.35$\%. Every structure have been optimized to calculate the optimum out-of-plane lattice parameter and by minimizing the forces down to 0.01~eV~\AA$^{-1}$. In these calculations, we employed an energy of 550~eV for the plane-wave expansion cutoff and the first Brillouin zone was sampled with a $6\times6\times1$ $\Gamma$-centered grid.

\subsection{Switching of the electric polarization}

In order to assess the role of the oxygen displacements through the switching of the electric polarization, we calculated the electronic structure for 7 different structures: 
\begin{itemize}
\item The physical properties for an electric polarization $P = \pm P_0$ are those calculated when the structure shown Fig.~1(a) is fully optimized.
\item   The structure with $P = 0$ was generated by placing all O1 and O2 atoms at the same $z$ position (along the [001] out-of-plane axis) and in the middle between two Hf layers; such transformation would correspond for the bulk HfO$_2$ to a displacement of the O1 atoms into the $z \simeq 0.375$ and $z \simeq 0.875$ positions, as shown in Fig.~\ref{figS1} described later, leading to a centrosymmetric structure with the space group $Pcca$ (No 54). With this slab, the positions of all inner Hf and O atoms were kept fix, while the interface oxygen atoms and all the Ni atoms were allowed to move during the structural optimization. 
\item  For the sake of simplicity, the structures with $P > 0$ and $P < \vert P_0 \vert$ have been built from a linear interpolation between the atom coordinates and lattice parameters of the two structures with $P = 0$ and $P = \pm P_0$, with no further structural optimization, explaining the linear variations of the Ni-O distances given in Fig.~3(a). The linear change of Hf/O coordinates can be considered as a linear variation of the electric polarization inside the HfO$_2$ film, if we make the approximation that the Born effective charges remain constant.
\end{itemize}
The results obtained from these calculations are reported in Fig.~3,~\ref{figS2} and~\ref{figS7}. Such calculations mostly intend at showing the importance of the Ni-O chemical bonds at the interface, rather than calculating precise variations of the different properties through a realistic switching of the electric polarization. Switching the electric polarization in HfO$_2$ can indeed occur through several pathways and would involve the movements of domain walls~\cite{Choe:2021,Ma:2023,Zhu:2024}, the study of which being out of the scope of this paper. 

\subsection{Magnetocrystalline anisotropy}

We performed non-self-consistent calculations to evaluate the magnetocrystalline anisotropy energy (MCAE) from the difference of total energies obtained for different orientation of the magnetization axis. These calculations were made with the VASP code in which the the spin-orbit interaction is included in the PAW spheres~\cite{Steiner:2016}. The projection of the MCAE $\Delta E^\textrm{SOC}_l$ on (001) atomic layers was computed by summing the contribution of every atom which forms layers $l$:
\begin{equation}
\Delta E^\textrm{SOC}_l = \frac{1}{A}  \sum_{atom~\in~l}E_{[0,0,1]}-E_{[i,j,0]} 
\end{equation}
($i,j$) being crystallographic indices, and $A$ the (001) surface of HfO$_2$. 

\subsection{Spin and orbital accumulations}

The out-of-equilibrium transport calculations use a full \textit{ab initio} DFT Hamiltonian matrices obtained directly from the \textsc{SIESTA} code~\cite{siesta_method}, employing atom-centered double-$\zeta$ plus polarization (DZP) basis sets. The calculations of these transport properties were performed using the atomic structures optimized with the VASP code. 
We used an energy cutoff for real-space mesh of \SI{400}{Ry} and a $\bm k$-mesh to sample the first Brillouin zone of $7\times7\times1$. The self-consistent spin-orbit coupling was introduced via the full off-site approximation~\cite{siesta_on-site_soc} using fully-relativistic norm-conserving pseudopotentials~\cite{tm_pseudopotentials}. The system Hamiltonian and overlap matrices were obtained after performing a full self-consistent cycle and treated within a post-processing routine based on SISL as an interface tool~\cite{zerothi_sisl}. This approach follows the same methodology as in our recent works~\cite{ovalle_2023,Liu2024,Krishnia:2024}.

The orbital or spin mean values, i.e. orbital and spin textures were calculated using: 

\begin{equation}
    \braket{\mathbf{\hat{O}}}_n=\braket{\psi_n|\hat{S}\mathbf{\hat{O}}|\psi_n},
\end{equation}

where $\ket{\psi_n}$ is an eigenstate with eigenvalue $\varepsilon_n$ of the hamiltonian $\hat{H}$. The overlap matrix $\hat{S}$ has to be considered because the basis is non-orthogonal. The matrices $\hat{O}_y = \hat{S}_y,\hat{L}_y$ are the usual spin and orbital angular momentum operator matrices expressed in terms of the atomic orbital basis, that is $p$ and $d$ orbitals. Finally, the spin and orbital accumulations were calculated using the Kubo formula:

\begin{equation}
   \delta \braket{\hat{O}_y} =   -e E_x \tau\int_{BZ} \partial_\epsilon f(\epsilon)d\epsilon \operatorname{Re}\left[ \braket{\psi_n|\hat{S}\hat{O}_y|\psi_n}\braket{\psi_n|v_x|\psi_n}\right]
   \label{eq:lresponse}
\end{equation}

where $v_x$ is the velocity operator modified for treating a non-orthogonal basis as

\begin{equation}
    \hat{v}_x= \frac{\partial 
    \hat{H}}{\partial k_x}
    -\frac{\varepsilon_n \partial 
    \hat{S}}{\partial k_x},
\end{equation}

$v_x$ is in the $\hat{x}$ in-plane direction along the external perturbation induced by an electric field $E_x$ of $10^4$~V~m$^{-1}$. The expression of the orbital or spin accumulation depends on the Fermi-distribution function $f$ and the momentum relaxation time $\tau$, the value of which having been chosen to correspond to the an energy broadening $\hbar \tau^{-1}= 75$~meV.

The contribution to the spin and orbital responses of each interface was obtained by constructing $\hat{O}_y = \hat{S}_y,\hat{L}_y$ in a block diagonal form such that

\begin{equation}
\hat{O}_y = \begin{pmatrix}
0& 0 & 0 & \hdots & 0 \\
0 & 0& \hdots & \hdots &\vdots \\
 0 & \vdots & \hat{O}_y^{l}&\vdots &0 \\
 \vdots & \hdots & \hdots &  0 & 0 \\
0 & \hdots & 0 & 0 & 0 \\
\end{pmatrix},\label{eq:big_hamiltonian}
\end{equation}

where $\hat{O}_y^{l}$ is the operator of the l-th layer.

\section{Atomic and electronic structure}

HfO$_2$ crystals can adopt different polymorphs~\cite{Huan:2014,Qi:2020,Zhu:2024}. Besides the high-temperatures cubic ($c$) and tetragonal ($t$) phases, hafnia can be found in a non-centrosymmetric and monoclinic ($m$) phase, with space group $P2_1/c$~\cite{Wang:1992}.  Among the two possible non-centrosymmetric orthorhombic structures, the $Pca2_1$ phase ($o$-phase) was confirmed by using aberration-corrected scanning transmission electron microscopy as responsible for the emergence of the ferroelectricity~\cite{Sang:2015}. The stabilization of this ferroelectric phase can be influenced by several physical parameters and, using first-principles calculations, Dogan, \textit{et al.} suggested that, in addition to the doping, the presence of an out-of-plane confinement, which can originate from the presence of electrodes, was necessary to preferably obtain the $o$-phase over the $m$-phase~\cite{Dogan:2019}. 

\begin{figure}[!htb]
\centering
\includegraphics[width=0.75\linewidth]{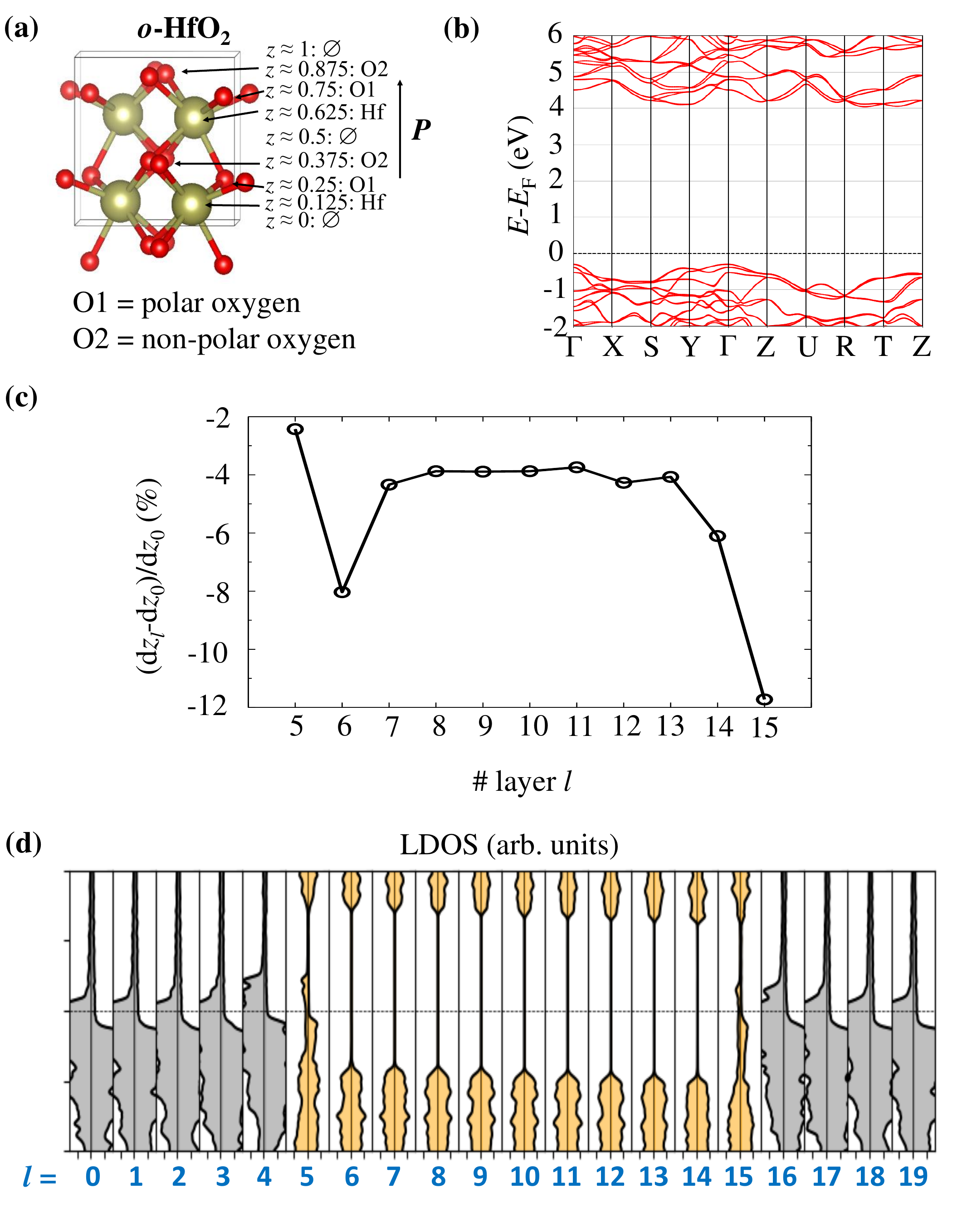}
    \caption{a) Schematic representation of the bulk HfO$_2$ atomic structure in its orthorhombic phase. The relative $z$ reduced coordinates of every atom are given approximately to differentiate the different atomic layers; the direction of the electric polarization can be switched by displacing O1 oxygen atoms between the $z\simeq0$/$z\simeq0.5$ and $z\simeq0.25$/$z\simeq0.75$ positions. (b) Electronic band structure calculated with the spin-orbit interaction. (c) Layer-resolved densities of states (LDOS) calculated for the complete Ni/HfO$_2$ superlattice. (d) Difference of cation-oxygen out-of-plane averaged coordinates $\textrm{d}z$ calculated in the different HfO$_2$ atomic layers and given as variations according to their bulk value $\textrm{d}z_0$. \label{figS1}}
\end{figure}

\subsection{Bulk structures}

\paragraph{HfO$_2$:} As shown in Fig.~\ref{figS1}(a), the orthorhombic ($o$-) phase of HfO$_2$, with space group $Pca2_1$ (No. 29), possesses two different types of oxygen atoms, O1 and O2, which are located inside polar layers or non-polar spacer layers, respectively. The calculated lattice parameters are $a = 5.211$~\AA, $b = 5.006$~\AA{} and $c = 5.028$~\AA, which is in good agreement with the literature~\cite{Muller:2012,hoffmann:2015,Sang:2015,Tao:2017}. The electric polarization is oriented along the [001] axis and has been calculated to $P_0^\textrm{bulk}=54.44$~$\mu$C~cm$^{-2}$ and 56.07~$\mu$C~cm$^{-2}$, using the methods based on the Berry phase or the Born effective charge, respectively; these values are also consistent with those reported in the literature~\cite{Huan:2014,Yang:2019,Zhu:2024}.

The band structure is shown in Fig.~\ref{figS1}(b). We calculated an indirect band gap of approximately 4.4~eV; the band-gap edges being shifted away from the $\Gamma$ and $T$ points because of the spin-orbit interaction, in agreement with the calculations performed by Tao~\textit{et al.}~\cite{Tao:2017}.

\paragraph{Ni:} We calculated a spin magnetic moment of 0.645~$\mu_\textrm{B}$ per Ni atom from the equilibrium fcc structure of bulk Ni. This value is decreased to 0.635~$\mu_\textrm{B}$ when the bulk Ni is strained by the in-plane lattice parameters of HfO$_2$. Regarding the magnetocrystalline anisotropy energy (MCAE), it is calculated to be very low for the unstrained bulk structure ($\simeq -5$~$10^{-2}$~meV~nm$^{-2}$). For the bulk Ni structure under strain, the MCAE decreases to  $-1.2$ $-1.5$ and $-1.8$~meV~nm$^{-2}$ between the out-of-plane [001] and [010], [110] and [100]  in-plane directions, respectively.

\subsection{Ni/HfO$_2$ superlattice}

Every atomic layer forming the HfO$_2$ layer depicted in Fig.~1(a) are formed by 2 formula units (f.u.) of HfO$_2$. As explained before, it is possible to distinguish two distinct types of oxygen atoms in the orthorhombic phase of HfO$_2$. As for the bulk compound, we labelled O2 the oxygen atoms which are approximately located in the middle between two Hf layers, and O1 the oxygen atoms which are off-centered and the position of which will mostly be responsible for the emergence of the out-of-plane electric polarization $\bm{P}$. 

As it can be seen in Fig.~\ref{figS1}(c), the layer-resolved densities of states (LDOS) display a linear shift in energy inside the HfO$_2$ film, characteristic of the presence of an internal electric field, which can be estimated around $1.73\times10^{8}$~V~m$^{-1}$. 

In Fig.~\ref{figS1}(d), we plotted the variations of the difference of Hf-O averaged coordinates along the out-of-plane [001] axis $\textrm{d}z$ and as a function of the atomic layers according to the two interfaces. $\textrm{d}z$ is defined as the difference between the averaged coordinates of the Hf atoms of a specific atomic layer $l$ and the averaged coordinates of the oxygen atoms O1 and O2 of the adjacent layers: $\textrm{d}z_l = \langle z(\textrm{Hf})\rangle_l - \langle z(\textrm{O1,O2})\rangle_{l-1,l+1}$. Even if in-plane displacements add small contributions to the polarization and if the Born effective charges may vary throughout the HfO$_2$ film, in a first approximation, we can still consider that the variation of $\textrm{d}z_l$ is representative of how the electric polarization varies in the film. We can see that $\textrm{d}z_l$ is reduced by approximately 3.9\% at the center of the HfO$_2$ film, which can be attributed to the presence of a depolarizing field. An oscillation of $\textrm{d}z_l$ is observed in the two atomic layers in the vicinity of the interface IF1, with an increase of $\textrm{d}z_l$ up to $-2.4$\% in the interface layer. $\textrm{d}z_l$ is on the contrary strongly reduced, down to $-11.7$\%, at the IF2 interface. 

\begin{table}[!ht]
\caption{Averaged chemical bond lengths $d$ and difference of $z$ coordinates $dz$ between the Ni and O atoms, atomic spin and orbital magnetic moments, $m_s$ and $m_\textrm{orb}$ at the two interface IF1 and IF2. The last column gives the excess of electrons $n_e$ calculated as the difference between the Mulliken charges $q_\textrm{Mul}$ (calculated with SIESTA) and the atomic number $Z$. \label{tab:IFproperties}} 
\centering 
\footnotesize
\begin{tabular}{cccccccccccc}               
\hline
\hline
Interface &  \multicolumn{2}{c}{$d_\textrm{Ni-O}$} & \multicolumn{2}{c}{$\textrm{d}z(\textrm{Ni-O})$} & \multicolumn{3}{c}{$m_\textrm{s}$} &  $m_\textrm{orb}$ & \multicolumn{3}{c}{$q_\textrm{Mul}-Z$}\\ 
 & \multicolumn{2}{c}{(\AA)} & \multicolumn{2}{c}{(\AA)} & \multicolumn{3}{c}{$\mu_\textrm{B}$/atom} & $\mu_\textrm{B}$/atom & \multicolumn{3}{c}{$n_e$/atom} \\
 &  O1 & O2 O1 & O2 & Ni & O1 & O2 & Ni  & O1 & O2 & Ni \\
IF1 & 2.05 & 1.94 & 0.95 & 1.47 & 1.07 & 0.28 & 0.15 & 0.10 & 1.16 & 1.09 & $-0.71$ \\
IF2 & 2.06 & 1.88 & 1.67 & 1.35  & 0.85 & 0.11 & 0.12 & 0.06 & 1.05 & 1.11 & $-0.52$ \\
$\Delta(\textrm{IF2}-\textrm{IF1})$ & 0.01 & $-0.06$ & $0.72$ & $-0.12$ & $-0.22$ &  $-0.17$ & $-0.03$ & $-0.04$ & $-0.11$ & 0.02 & 0.19 \\
\hline
\hline
\end{tabular}
\end{table}

In addition to the Hf-O rumpling along the out-of-plane direction, it is important to analyze the Ni-O interatomic distances at the interfaces. As shown in Table~\ref{tab:IFproperties},  we can notice strong variations of the out-of-plane coordinates, which is a direct consequence of the reversal of the electric polarization and of the substitution of O1 or O2 atoms at the position of first neighbors of the Ni atoms. These variations of distances result in different orbital overlaps and charge transfer; the interface band hybridizations and charge transfer have for effect to induce an increase of the spin magnetic moment of the Ni atoms by 65\% and 31\% at the interface IF1 and IF2, respectively. Consecutively, a spin magnetic moment is induced on the oxygen atoms first-neighbor of the Ni electrode, which is maximal for the O1 atoms of IF1 which possess the lowest $\textrm{d}z(\textrm{Ni-O})$ value. The difference of magnetic moment between the two interfaces is characteristic of a chemical-bond mechanism responsible for a magnetoelectric coupling, with a total magnetic-moment difference of 1.44~$\mu_\textrm{B}$ for an interface containing 4 Ni and 4 O atom, in agreement with the value of 1.49~$\mu_\textrm{B}$ computed by Yang, \textit{et al.}~\cite{Yang:2019}.

\begin{figure}[!htb]
    \centering
    \includegraphics[width=1\linewidth]{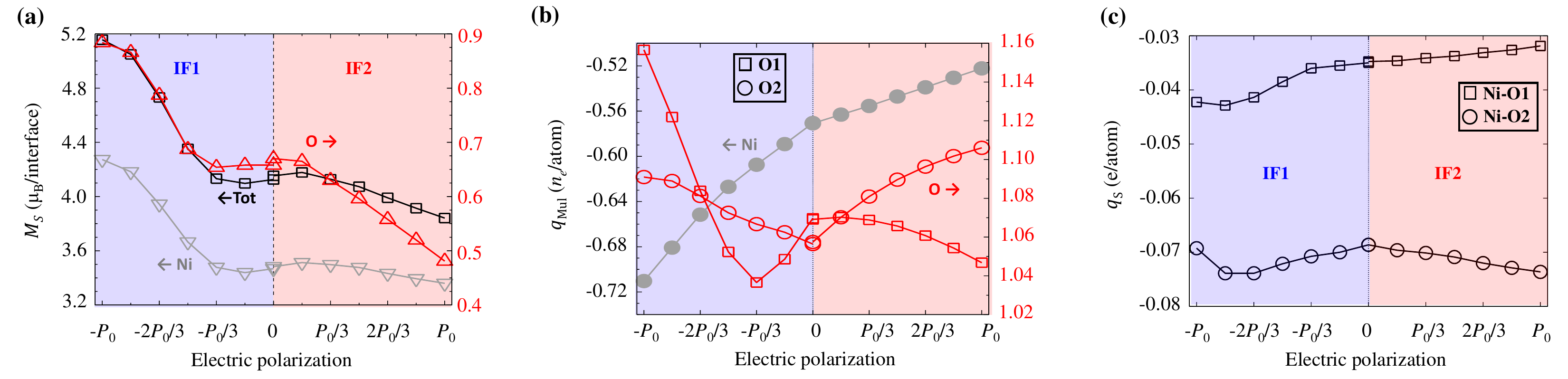}
    \caption{Variations as a function of the electric polarization of (a) the spin magnetic moment per interfaces of 4 Ni and 4 O atoms (calculated using VASP), (b) the Mulliken charges in excess of electrons per atom (calculated using SIESTA) and (c) the overlap Mulliken charges between Ni and O atoms (calculated using SIESTA). \label{figS2}}
\end{figure}

In Fig.~\ref{figS2}(a), we can see that the variations of spin magnetic moments is not linear when the electric polarization is linearly varied from $-P_0$ to $+P_0$. These non-linear and non-monotonous changes follow the variations of Ni-O chemical-bond lengths given in Fig.~3(a) and are consistent with the variations of charge transfer between Ni and O1 given in Fig.~\ref{figS2}(b), the variations of overlap Mulliken charges of Fig.~\ref{figS2}(c), but also with the variations of the $d$-$d$ interactions which govern the magnetocrystalline anisotropy Fig.~3(b). Fig.~\ref{figS2} finally demonstrates the good match between the calculations performed with the VASP and SIESTA codes.  

\begin{figure}[!htb]
    \centering
    \includegraphics[width=1\linewidth]{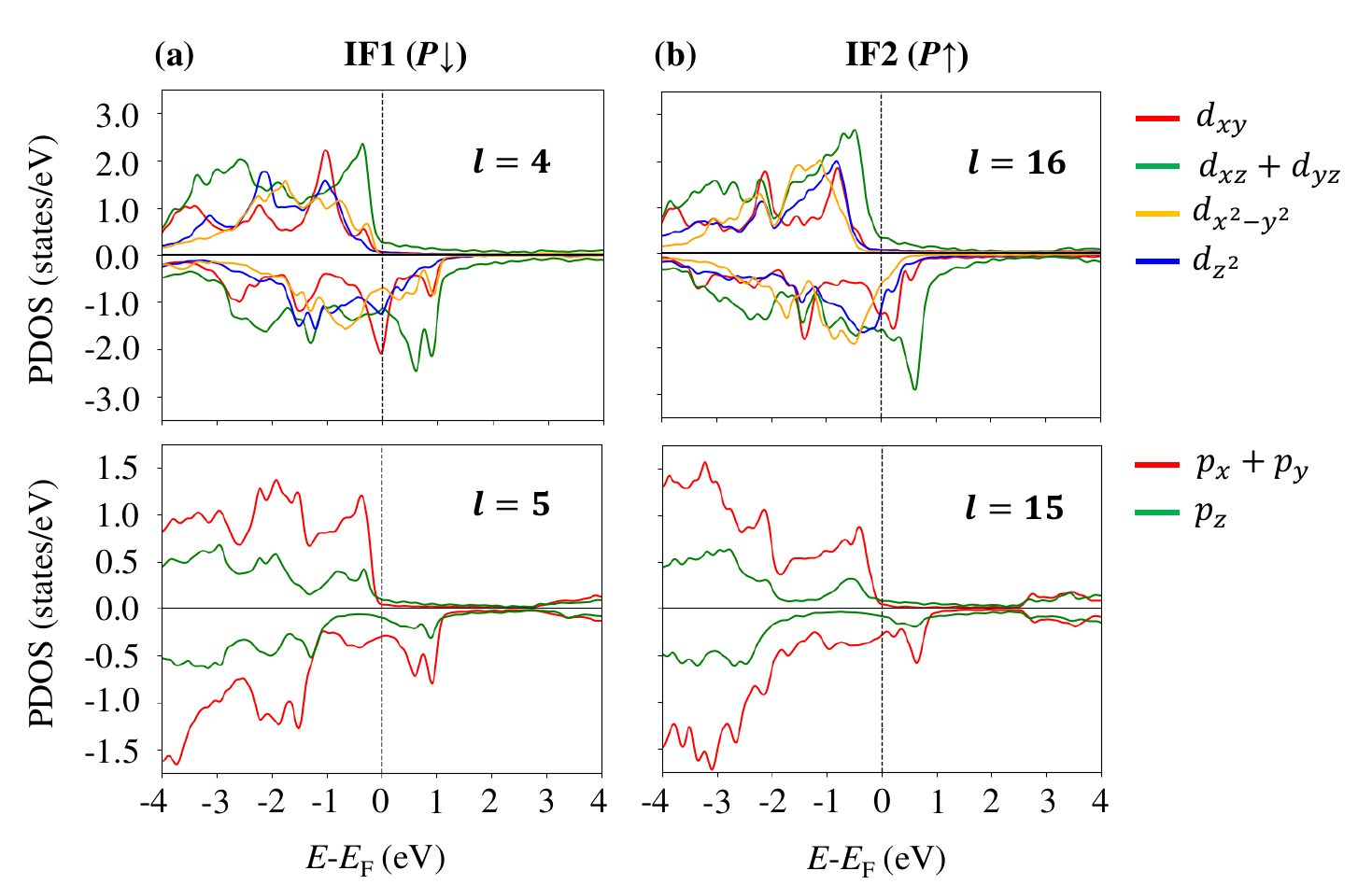}
    \caption{Orbital-projected densities of states (PDOS) calculated at the two interfaces IF1 and IF2.\label{figS3}}
\end{figure}

In Fig.~\ref{figS3}, we report the orbital-projected densities of states (PDOS) associated with the contributions of the Ni and O atoms at the two interfaces. The $x$, $y$ and $z$ axes are defined according to the cubic lattice of the Ni atoms. The PDOS analysis is complex due to the large number of bands near the Fermi level and the strong hybridization between Ni-$d$ and O-$p$ bands. Because of the presence of the interface and of the tensilfe in-plane strain, we can see that the peaks of DOS associated to the in-plane $d_{xy}$ orbitals are stabilized and appear at lower energies than the peaks with a$d_{xz+yz}$ character. For the interface IF1, we can notice a net peak of DOS with a $d_{xy}$ character at the Fermi level, for the minority-spin channel. At IF2, this peak is slightly shifted above the Fermi level, where $d_{xz+yz}$+O-$(p_x,py)$ unoccupied bands contribute dominantly to the DOS.

\section{Spin and orbital accumulations \label{ap:sec:Layer}}

\begin{figure}[!htb]
    \centering
    \includegraphics[width=1\linewidth]{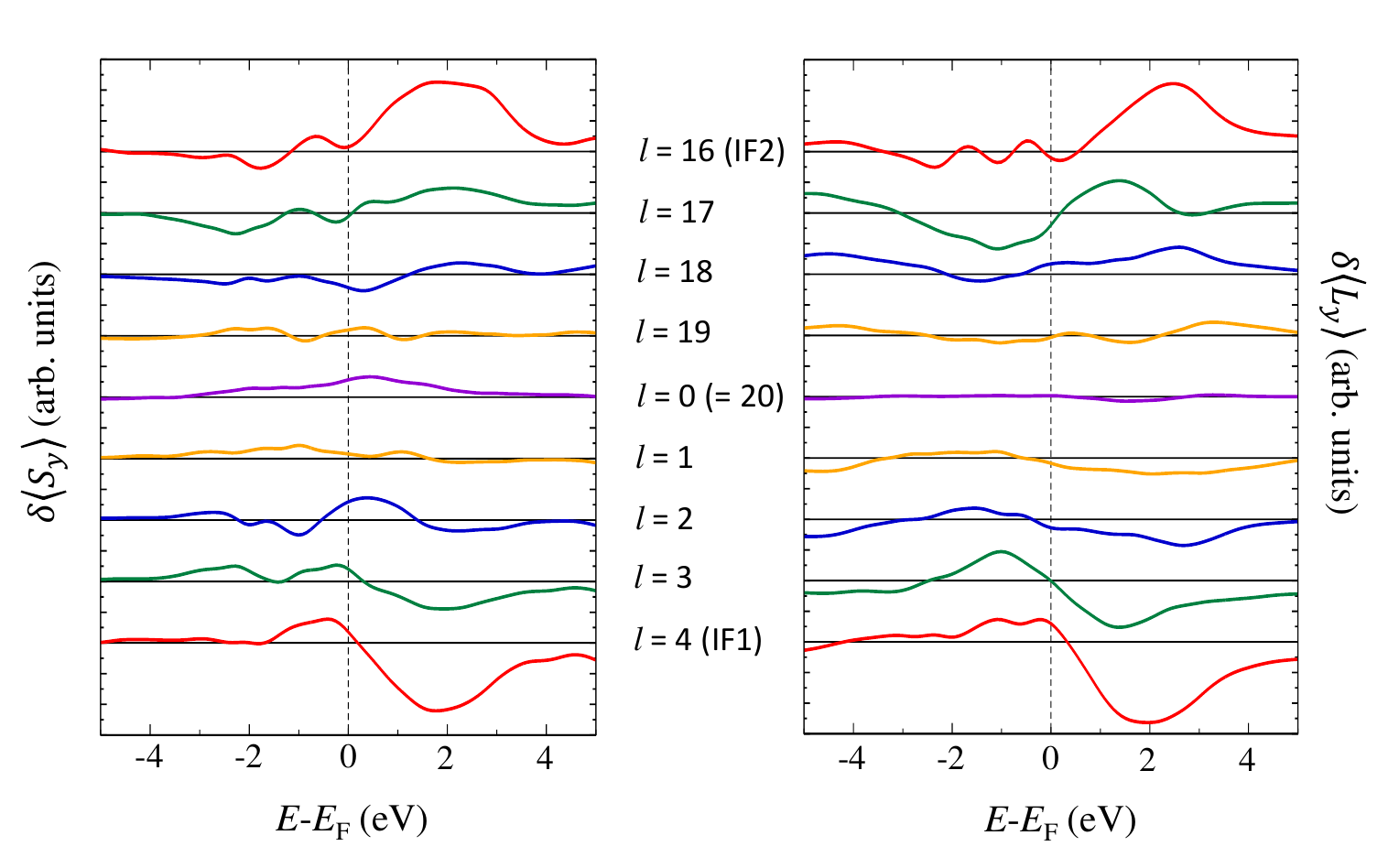}
    \caption{Layer-resolved spin ($h_{S_y}$) and orbital ($h_{L_y}$) response calculated for the two different interfaces. \label{figS4}}
\end{figure}

\begin{figure}[!ht]
    \centering
    \includegraphics[width=\linewidth]{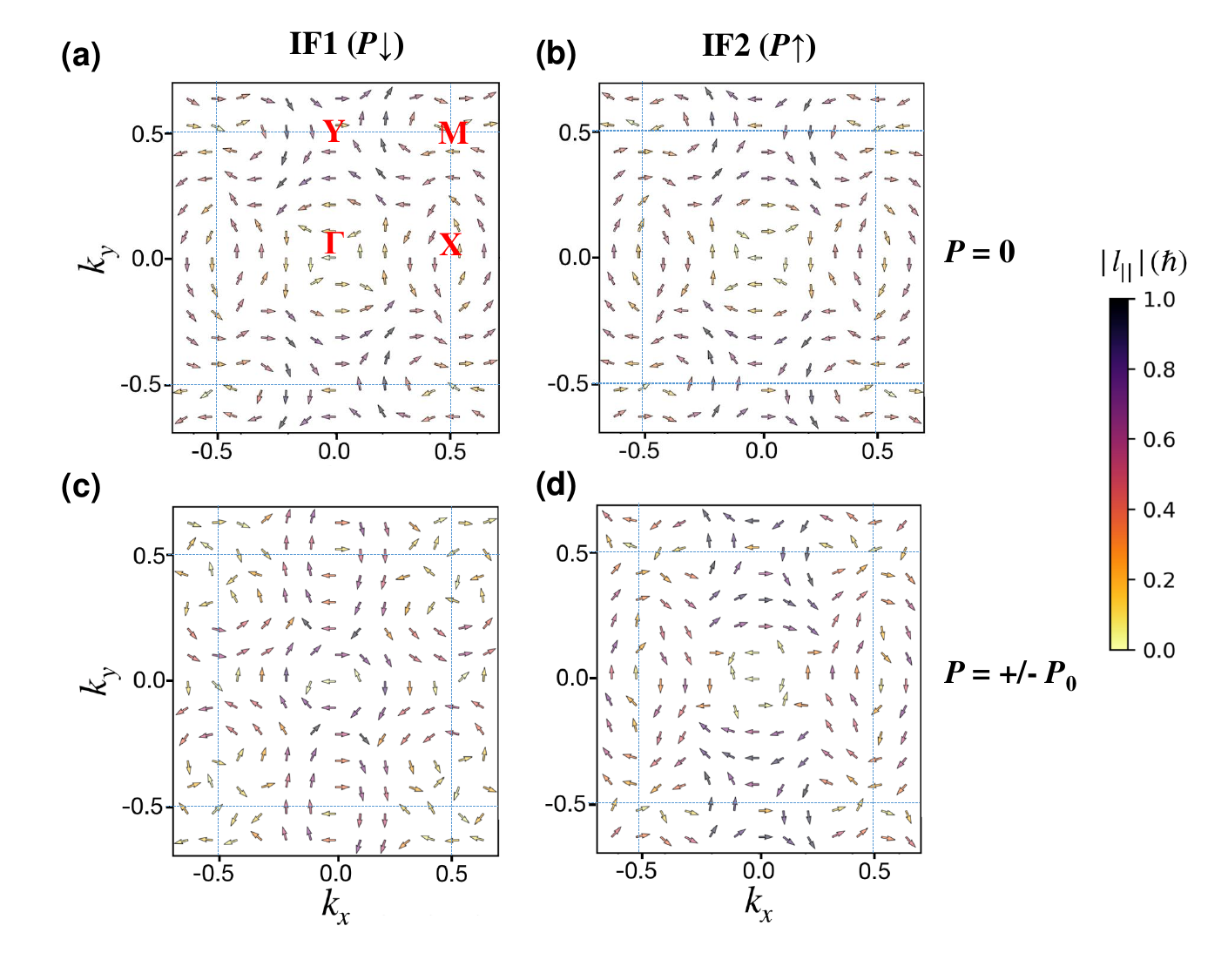}
    \caption{Orbital textures calculated with the superlattices with (a,b) $P = 0$ and (c,d) $P = \pm P_0$ and at the energy $E = E_\textrm{F}$. The color bar represents the magnitude of the in-plane component of the orbital texture. Because of the lattice distortions resulting from the orthorhombic structure of HfO$_2$, the orbital textures are essentially non zero along the $\Gamma-Y$ direction. For $P = 0$, we clearly see a change in the chirality when comparing the calculated texture at IF1 and IF2. For $P = \pm P_0$, we note a change in the shape of the texture, switching from a Rashba-like texture at IF1 to a Dresselhaus-like shape  at IF2. \label{figS5}}
\end{figure}

As shown in Fig.~\ref{figS4}, the strongest contributions to the spin and orbital responses to the electric field come from the interface layers [$l=4$ (IF1) and $l=16$ (IF2)]. The layer-resolved responses decrease down to approximately a zero value 3 atomic layers further ($l=1$/$l=19$). Such a decrease is certainly mostly due to the decrease of the interface perturbation in terms of charge transfer and lattice distortion but it may also partly be due to the odd function of the responses $\delta \braket{\hat{S}_y/\hat{L}_y}$ as a function of the out-of-plane $z$ coordinate. Indeed, it is important to note that the orbital and spin textures are chiral (see Fig.~\ref{figS5}): we indeed built our superlattices such that IF1 can be viewed as an interface with a bottom electrode and IF2 with a top electrode, in other words, such that the $z$ axis is inverted between the two interfaces, resulting in an opposite sign for the responses calculated for these two interfaces. In the main paper, because our goal was to compare the effect of the switching of the electric polarization only, the responses of the interface IF1 have systematically been multiplied by $-1$.

\begin{figure}[!ht]
    \centering
    \includegraphics[width=\linewidth]{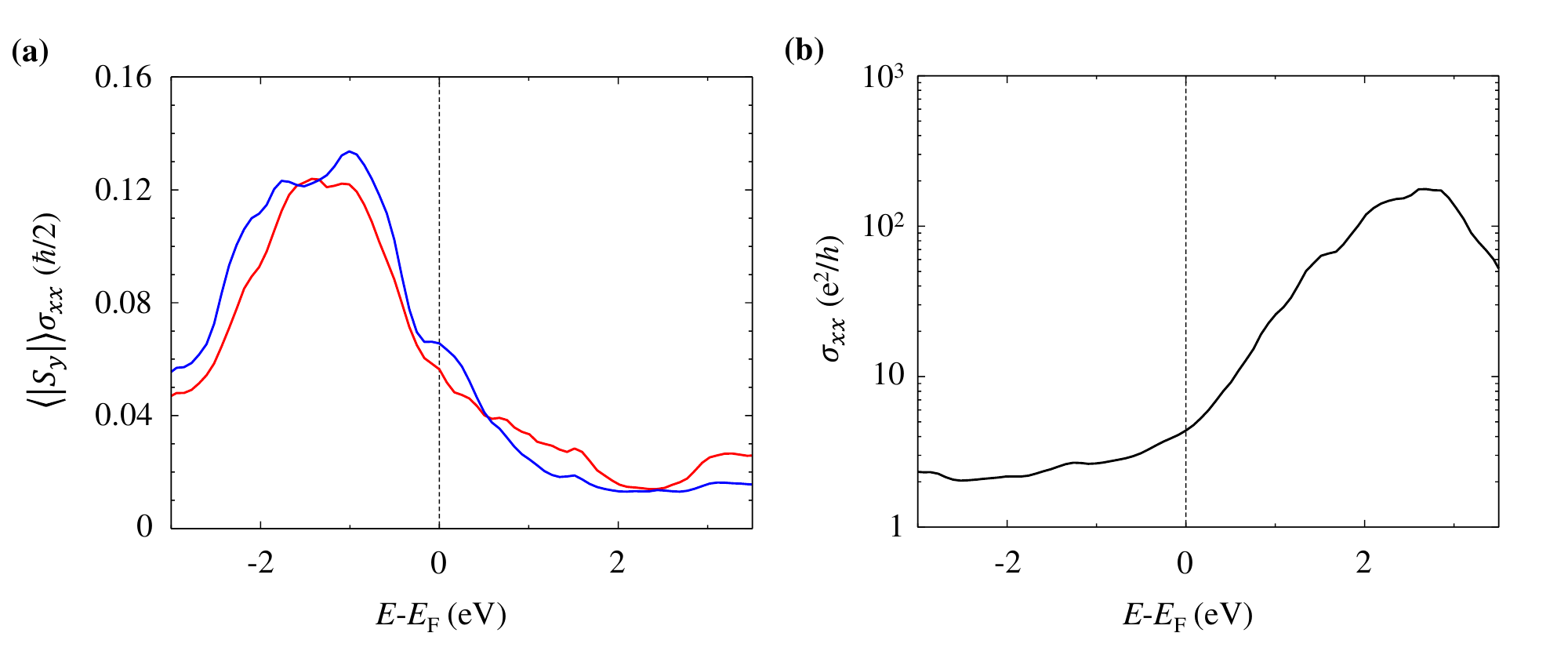}
    \caption{(a) Spin polarization of Ni and (b) longitudinal conductivity $\sigma_{xx}$ as a function of the chemical potential.}
    \label{figS6}
\end{figure}

Finally, in Eq.~\ref{eq:lresponse}, we remind that the spin and orbital accumulations are functions of an integration over the Brillouin zone of a product of the mean operator values $\braket{\hat{O}}$ times the velocity $\braket{v_x}$. In Fig.~\ref{figS6}, we show that the variation of spin/orbital accumulations will mostly depend on:
\begin{itemize}
    \item the variation of the first term of the product, which reaches values as high as $\braket{\hat{S}_y} = 0.20 \hbar$ for energies below the Fermi level, where the transport is dominated by Ni-$d$ bands. Above the Fermi level, this value is vanishingly small as it will be associated to $sp$ states.
    \item the variation of the velocity $v_x$ is accounted by considering the longitudinal conductivity $\sigma_{xx}$ $\propto$ $\int_{BZ} \partial_\epsilon f(\epsilon)d\epsilon \operatorname{Re}\left[ \braket{\psi_n|\hat{v}_x|\psi_n}\braket{\psi_n|\hat{v}_x|\psi_n}\right]$ which displays a large value above the Fermi level, increasing by 2 orders of magnitude compared to its value for the filled states because of the dispersion of $sp$ bands which is larger than the dispersion of $d$ bands.  
\end{itemize} 

\begin{figure}[!ht]
    \centering
    \includegraphics[width=\linewidth]{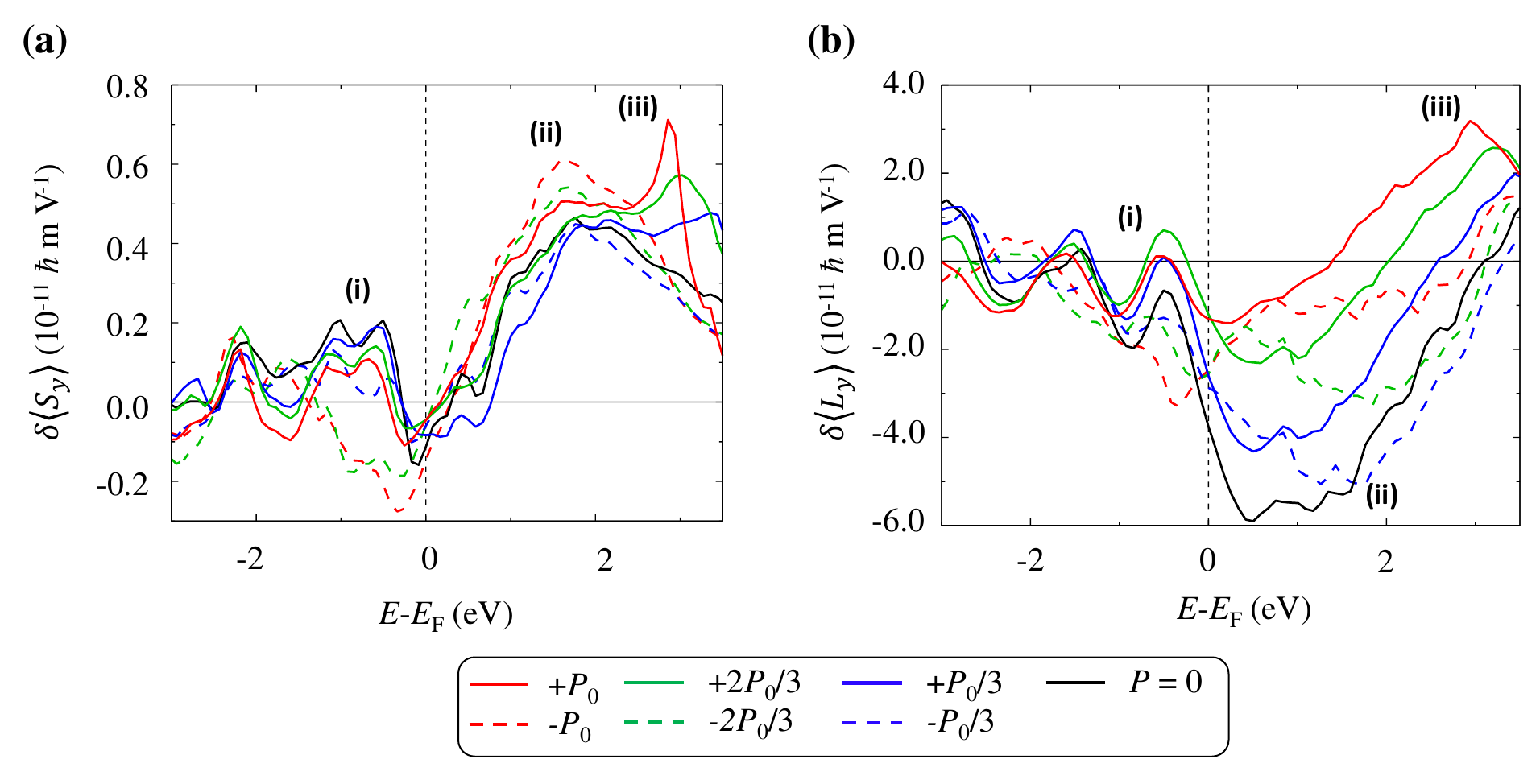}
    \caption{Variation of the total (a) spin and (b) orbital accumulations as a function of the electric polarization. \label{figS7}}
\end{figure}

The variation of the spin and orbital responses as a function of the polarization direction and magnitude are shown in Fig.~\ref{figS7}. 

\end{document}